\documentclass[12pt]{iopart}

\usepackage{iopams}
\usepackage{graphicx}
\usepackage{dcolumn}
\usepackage{times,mathptm}
\newcommand{\beq}{\begin{equation}}
\newcommand{\eeq}{\end{equation}}
\newcommand{\bea}{\begin{eqnarray}}
\newcommand{\eea}{\end{eqnarray}}

\renewcommand{\d}{\delta}

\renewcommand{\L}{\Lambda}
\renewcommand{\b}{\beta}
\renewcommand{\a}{\alpha}

\renewcommand{\k}{\kappa}

\newcommand{\n}{\nu}
\newcommand{\m}{\mu}
\renewcommand{\r}{\rho}
\renewcommand{\P}{\Phi}

\newcommand{\s}{\sigma}

\renewcommand{\S}{{\cal S}}

\renewcommand{\e}{\epsilon}

\newcommand{\oh}{{\textstyle{\frac{1}{2}}}}

\newcommand{\dg}{\dagger}
\newcommand{\non}{\nonumber}

\newcommand{\rf}[1]{(\ref{#1})}
\newcommand{\ra}{\rightarrow}
\newcommand{\pa}{\partial}

\begin{document}
%
% Front matter
%
\title{Quark Confinement: The Hard Problem of Hadron Physics}

\author{R. Alkofer}
\address{Institut f{\"ur} Physik, Karl-Franzens-Universit{\"a}t, Universit{\"a}tsplatz 5,
A-8010 Graz, Austria}
\author{J. Greensite}
\address{Physics and Astronomy Department, San Francisco State
University, San Francisco, CA~94132, USA}
\date{\today}
\begin{abstract}

   We give a brief overview of the problem of quark confinement in hadronic
physics, and outline a few of the suggested explanations of the confining force.

\end{abstract}

%
%\keywords{Confinement, Lattice Gauge Field Theories}
%
%\maketitle
%
% Section I
%
\section{Introduction}\label{Introduction}

     The hadron spectrum found in nature consists of color singlet combinations of color non-singlet objects:
the quarks and gluons.   Unlike atomic physics, where electrons can readily be separated from atoms,
there is no color-charge version of ionization in hadronic physics.
Every attempt to kick a quark free from a hadron, via high-energy collisions, only results in the production
of more color-singlet hadrons; a non-singlet particle is never produced.   
Particle and nuclear physicists have become accustomed to this fact,
which is often referred to as ``color confinement", but after thirty-four years of intense effort this
very  basic feature of hadron physics still has no generally agreed upon explanation.   Color confinement is
therefore a hard problem.   In this article we would like to discuss some aspects of this problem which we
think are important, and to briefly survey a few of the main avenues of research. 
 
     To begin with, what would be the energy of an isolated quark?  In gauge theories, abelian or non-abelian,
a charge density, $\rho_{\mathrm quark}^a$, is the source of a longitudinal electric field, as required by the Gauss Law
\beq
    \vec{\nabla} \cdot \vec{E}^{a} =  \rho_{\mathrm quark}^a - gf^{abc}A^b_k E^c_k
\eeq
where the term containing the structure constant of the gauge group $f^{abc}$ and the gauge field $\vec{A}^{a}$ 
is only present in non-abelian gauge theories and reflects the non-vanishing color electric charge of the gluons.
Their charge is in the {\bf 8} representation of the SU(3) gauge
group, and cannot neutralize the color charge of a quark in the {\bf 3} representation.  So the color electric field
of an isolated quark could only end on another isolated quark, or else extend out to infinity.
The fact that isolated quarks are not seen in nature means that the energy stored in the associated color electric field
must be very large.  But \emph{how} large?  Suppose we try to free a quark from a hadron by hitting it with a
high energy (real or virtual) photon.   As the struck quark begins to move away from the other quarks in the hadron,
it brings along the color electric field necessary to satisy the non-abelian Gauss Law.   If the energy stored in the color
electric field becomes large enough, then the system is unstable to light quark-antiquark pair creation.  The 
antiquark of the pair binds to the struck quark, resulting in a color singlet, and the quark of the pair binds to the
remaining quarks of the hadron, forming another color singlet.  The two color singlet hadrons are generally still
in highly excited states, and decay into lighter hadrons.  The end result is a shower of ordinary hadrons, rather than
a free quark and a color-ionized hadron.

     So a hadron scattering experiment will not answer our question about the energy stored in the color electric field
of a free quark, at least not directly.   Our knowledge about this energy is therefore indirect, and comes from two sources:
numerical simulations, and a pattern in the hadron spectrum known as Regge trajectories.  Let us imagine ``dialing" 
the bare quark mass parameters in the QCD Lagrangian
so that all quarks are very heavy; so heavy, in fact, that pair creation processes do not become important until 
quark separations reach macroscopic (or even cosmic!) distances.   Then, starting with a tightly bound color singlet object such as 
a meson, and measuring the energy required to slowly separate the massive constituent quark and antiquark by a distance $R$, we 
get an estimate for the static quark potential $V(R)$, and this is essentially a measure of the energy stored in the
color electric field due to the quarks.  Of course, in nature the current quark masses are whatever they are, and cannot be
changed, but on a computer anything is possible: Nothing prevents us from simulating a version of QCD with very
massive quarks, as we will discuss in more detail below.  

    Our second source of information about the static quark potential is derived from the actual hadron spectrum.  In the
spectrum there exist certain metastable states which are sufficiently long-lived to show up as resonances in scattering
cross-sections.  The fact that it takes some time for these metastable states to decay via quark pair creation means that,
for the short period prior to decay, the resonances are sensitive to interquark forces in the absence of liqht quark pair
creation.  From the masses of the resonant states,  we can therefore learn a great deal about states with comparatively large quark
separations, and about the energy which is stored in the associated color electric fields.

\section{The Linear Potential}
  
    The following theorem \cite{Bachas} can be proven in lattice gauge theory:  
the force between a static quark and
antiquark is always attractive but cannot increase with distance, {\it i.e.}
\beq
               {\pa V \over \pa R} > 0  \; , \quad {\pa^{2} V \over \pa R^{2}} \le 0 \; .
\eeq
The second inequality is saturated by a linear potential; the static quark potential can rise no faster 
than linearly with distance.  The theorem does not tell us that the static quark potential actually 
\emph{does} rise linearly, but hadron phenomenology suggests, and computer simulations convincingly 
demonstrate, that this is the true, or at least very close to the true, 
behavior of the potential at large quark separations.
 
\subsection{Regge Trajectories and the Spinning Stick Model}

\begin{figure}
\centering
\includegraphics[height=10cm]{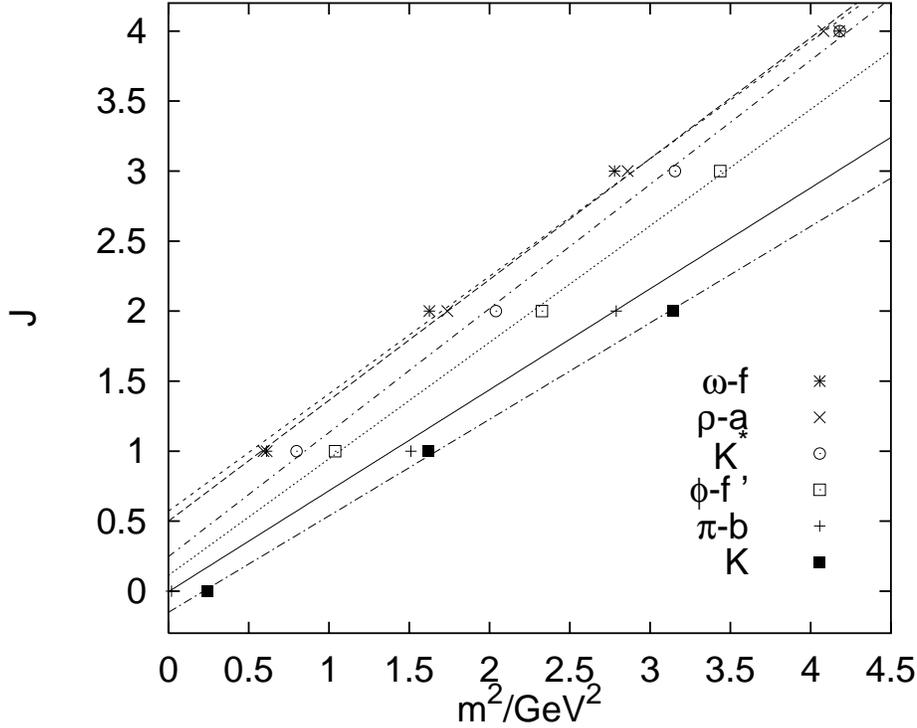}
\caption{Regge trajectories for the low-lying mesons (adapted from
 ref.\ \cite{Bali0}).}
\label{regge}     
\end{figure}

    A remarkable pattern emerges in the hadronic spectrum,  when the spin of mesons (and baryons) is plotted against
their squared mass, as shown in Fig.\ \ref{regge}.  In such plots the mesons and baryons of given flavor
quantum numbers seem to lie on nearly parallel straight lines, known as linear Regge trajectories.
This is a very striking feature of the hadronic spectrum, nothing similar is found in the electroweak
theory, and the question is why it occurs.

    Suppose that we picture a meson as a straight line of length $L=2R$, with mass per unit length
$\s$.  The line rotates about a perpendicular axis through its midpoint,  such that the endpoints of
the line are moving at the speed of light, $v(R)=c=1$.  Then for the energy in the rest frame,
 {\it i.e.\/} the mass, of the spinning stick we have
\bea
       m = {\rm Energy} = 2 \int_{0}^{R} {\s dr \over \sqrt{1 - v^2(r)}}
         =  2 \int_{0}^{R} {\s dr \over \sqrt{1 - r^2/R^2}}
         = \pi \s R \; ,
\label{mR}
\eea
and for the angular momentum
\bea
       J = 2 \int_{0}^{R} {\s r v(r)dr \over \sqrt{1 - v^2(r)}}
         = {2\over R} \int_{0}^{R} {\s r^2 dr \over \sqrt{1 - r^2/R^2}}
         = \oh \pi \s R^2 \; .
\eea
Comparing the two expressions, we see that
\beq
 J = {1 \over 2 \pi \s} m^2 = \alpha' m^2
\eeq
The constant $\a'$ is known as the \emph{``Regge slope"}.

     From the data one estimates $\a' = 1/(2\pi \s)= 0.9~ \mbox{GeV}^{-2}$, which gives a
mass/unit length of the string, or \emph{``string tension"}, of
\beq
\sigma \approx 0.18~ \mbox{GeV}^{2} \approx 0.9~ \mbox{GeV/fm} .
\eeq
The spinning stick model is, of course, only a caricature of the real situation. 
In fact the various Regge trajectories
do not pass through the origin, and have slightly different slopes.  To make
the model more realistic, one might want to relax the requirement of
rigidity, and allow the ``stick" to fluctuate in transverse directions.  This line
of thought leads to string theory.  However, since QCD is the
theory of quarks and gluons, the question to be answered is how
a stick-like or string-like object actually emerges from that theory.

    One possible answer is via the formation of a color electric flux tube.  We imagine
that the color electric field running between a static quark and antiquark is, for some
reason, squeezed into a cylindrical region, whose cross-sectional area is nearly constant
as quark-antiquark separation $L$ increases.  In that case,
the energy stored in the color electric field will grow linearly with quark separation, 
{\it i.e.\/}
\beq
\mbox{Energy} = \s L ~~~\mbox{with} ~~~
\sigma = \int d^2x_\perp ~\oh \vec{E}^a \cdot \vec{E}^a
\eeq
where the integration is over a cross-section of the flux tube.
This means that there will be a linearly rising potential energy associated with static
sources (the ``static quark potential"), and an infinite energy is required to separate these
charges an infinite distance.   

    In this way the pattern of metastable states in the hadron spectrum suggests a picture of
how the color electric field energy, in the absence of light quark pair creation, would grow 
with quark separation. 
 
\subsection{Wilson Loops and Lattice Simulations}

      The most reliable evidence we have about the static quark potential is obtained from computer
simulations of quantum chromodynamics.   For this purpose it is useful to simulate
a version of QCD in which the quarks are very massive, and pair creation in the vacuum can
be ignored.

      Let $Q(t)$ be the creation operator of a state at time $t$ containing a very massive quark and a very massive
antiquark, separated by a distance $R$.  There are many operators of that sort, but, unless we fix a
gauge, it is necessary for $Q$ to be gauge-invariant.  If not, then $Q$ and correlators of $Q$ will
simply average to zero in the functional integral over gauge fields $A$ and the quark fields $\psi$.  Consider
the unequal-times correlator
\bea
             \langle Q^{\dg}(T)  Q(0) \rangle = {1\over Z}\int DA D\psi D\overline{\psi} \;
                      Q^{\dg}(T)  Q(0)\; e^{iS}
            = \langle \Psi_{0} | Q^{\dg} e^{-i(H-{\cal E}_{0})T} Q |\Psi_{0}\rangle
\eea
where $H$ is the Hamiltonian operator, ${\cal E}_{0}$ is the vacuum energy and $\Psi_{0}$ is the vacuum state, 
in any gauge (the gauge choice
does not matter if $Q$ is gauge invariant).   By transforming the theory from Minkowski space to Euclidean
space by a Wick rotation  of the time coordinate $t\ra it$, and inserting a complete set of energy eigenstates
$\{\Psi_{n}\}$ with the quantum
numbers of the heavy quark-antiquark pair, the above expression becomes
\goodbreak
\bea
        \langle Q^{\dg}(T)  Q(0) \rangle_{\rm euclid} &=& {1\over Z}\int DA D\psi D\overline{\psi} 
                      Q^{\dg}(T)  Q(0) e^{-S}
\non \\
                   &=& \langle \Psi_{0} | Q^{\dg} e^{-(H-{\cal E}_{0})T} Q |\Psi_{0}\rangle
\non \\
       &=& \sum_{n} \Bigl|\langle \Psi_{0}|Q^{\dg}|\Psi_{n} \rangle\Bigr|^{2} e^{-E_{n}T} 
\eea
where $E_{n}$ is the excitation energy (above ${\cal E}_{0}$) of the energy eigenstate $\Psi_{n}$.  
The notation $\langle \rangle_{\rm euclid}$ indicates Euclidean-time expectation values, but from here on 
the ``euclid" subscript will be dropped. 
We see that at large Euclidean time separations, the correlator is dominated by the minimum energy state
of the Minkowski theory.   So to find the \emph{minimum} energy possible for a state containing a heavy quark
antiquark pair, satisfying the Gauss Law, we have to calculate
\beq
          E_{min}^{q\overline{q}}(R) = -\lim_{T\ra \infty} {d \over dT} \log\Bigl[ \langle Q^{\dg}(T)  Q(0) \rangle \Bigr]
\; .
\eeq 
At large $T$ any choice of $Q$ will do, providing $Q$ is gauge-invariant, and the quarks are created at separation $R$.
The simplest choice is
\beq
            Q = \overline{\psi}(0) \; P\exp[i\int_{0}^{R} dx^{\m} A_{\m}] \; \psi(R)
\eeq
where the expression between the quark operators is a path-ordered exponential of the matrix-valued A-field, which lies
in the Lie algebra of the gauge group.  If the quarks are very heavy and quark loops can be ignored, then the functional
integration over the quark fields can be carried out explicitly, with the result
\beq
        \langle Q^{\dg}(T)  Q(0) \rangle = \kappa M^{-2T} \left\langle 
              \mbox{Tr}\left[ P e^{i\oint_{C} dx^{\m} A_{\m}}\right] \right\rangle \; .
\eeq
In this expression $\k$ is a numerical prefactor, coming from a trace over Dirac matrices,  and $M$ is a constant
which depends on the bare quark mass (and the UV regulator).  Neither of these terms have any sensitivity to the quark
separation $R$.  The term we are interested in is the remaining expectation value, which involves only the 
gluon field
\beq
            W(C) =   \left\langle \mbox{Tr}\left[ P e^{i\oint_{C} dx^{\m} A_{\m}}\right] \right\rangle 
\eeq
where in this case the line integral runs around a closed rectangular contour $C$ of length $T$ and width $R$. 
The path-ordered exponential of such line integrals are known as \emph{Wilson loops}.  For a rectangular contour,
we will denote the expectation value by $W(R,T)$, 
and the part of the potential which depends on $R$ is extracted from 
\beq
             V(R) = -\lim_{T\ra \infty} {d\over dT} W(R,T)\; .
\eeq
From now on this will serve as our definition of the heavy quark potential.

      Nobody knows how to calculate $W(R,T)$ analytically in QCD, when $R$ and $T$ are large compared to the
length scale set by the fundamental scale of QCD, $\L_{\tt QCD}$.   
However, this quantity can be calculated in regions of interest by numerical
simulations, which require regulating QCD on a finite lattice.  Lattice gauge theory is explained in detail in a number
of texts, {\it e.g.\/} see ref.\ \cite{Smit}, but very briefly the idea is this:  Continuous spacetime is replaced by a
D=4 dimensional hypercubic lattice; the points on the lattice are called ``sites", the lines between neighboring sites
are ``links", and the squares bounded by four neighboring links are called ``plaquettes".   The Lie algebra-valued
gauge field $A_{\m}(x)$ of the continuum theory is replaced by a set of group-valued link variables $U_{\m}(x)$, 
associated with links of the lattice.  A Wilson loop is simply the trace of the product of link variables around a closed contour
on the lattice, with the understanding that when the contour runs through a link in the negative x, y, z, or t directions,
then the hermitian conjugate $U^{\dg}_{\m}(x)$ is used in place of $U_{\m}(x)$.  The lattice version of the 
functional integral over gauge fields,
\beq
            Z = \int DU ~ e^{S_{W}} \; , 
\eeq 
is based on the ``Wilson action" $S_{W}$
\beq
            S_{W} = {\b \over 3}\sum_{x}\sum_{\m=1}^{3}\sum_{\n > \m} \mbox{Tr}
            \Bigl[ U_{\m}(x) U_{\n}(x+\m)U_{\m}^{\dg}(x+\n)U^{\dg}_{\n}(x)\Bigr] + \mbox{~c.c} \; .
\eeq
Numerical (``lattice Monte Carlo") simulations involve stochastically generating a set of lattice gauge field 
configurations according to the 
probability distribution $P[U] = e^{S_{W}}/Z$; an estimate of the vacuum expectation value of some operator (such as
a Wilson loop) is simply the average value that the operator takes in the set.   Every quantity calculated on the lattice
is calculated in units in which the lattice spacing $a=1$.  To convert to physical units, $a$ must be assigned a value in,
say, fermis, and all lattice results are a function of the lattice spacing and the coupling $\b=2N/g^{2}$ for the gauge
group SU(N).  
Of course, the masses of hadrons should not depend on the lattice spacing, but renormalization theory 
teaches us that at sufficiently small couplings, a change in $a$ can be compensated for by a change in $g$, leaving 
the spectrum and other physical quantities invariant.

\begin{figure}
\centering
\includegraphics[height=10cm]{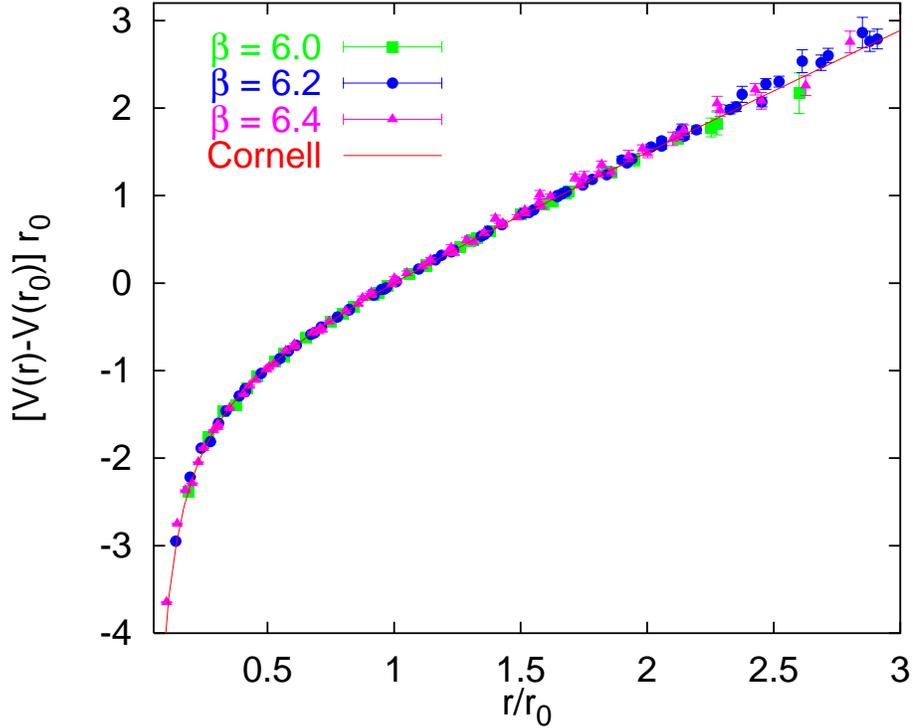}
\caption{The static quark potential, obtained from lattice Monte Carlo simulations,
in SU(3) gauge theory.  Both potential and quark separation are in units of the ``Sommer
scale" $r_0\approx 0.5 fm$ (adapted from ref.\ \cite{Bali0}). }
\label{pot}     
\end{figure}
    An example of the static quark potential obtained by lattice Monte Carlo techniques is shown in Fig.\ \ref{pot}.
Here the potential, computed at several $\b$ values, is plotted against quark separation in units of a certain physical scale 
$r_{0}$, which is about 0.5 fm.   In this graph we see very convincing confirmation of the linearity of the static quark
potential at large distances.   If the potential rises linearly indefinitely, then the energy of an isolated quark would be
infinite.  It is no wonder, then, that color non-singlet particles are not produced in hadronic collisions.

    The linear rise of the static quark potential at arbitrarily large quark separations is a rather
astonishing  and important fact, and the question $-$ the quark confinement problem $-$ is how to
account for such behavior.  Until that question is answered satisfactorily, we do not really
understand hadronic physics, nor do we understand the dynamics of non-abelian gauge theories at
large distance scales.

\subsection{Further Properties of the Linear Potential}

        In addition to varying the quark mass in numerical simulations, one can also vary the color group representation
of the ``quarks" (i.e.\ heavy static color sources), and study the effect on the static quark potential.  
Numerical simulations, and some general arguments, indicate that there are two distinct sorts of representation dependence, 
depending on the static source separation:
\begin{enumerate}
\item {\bf Casimir Scaling.}  Initially the slope of the linear potential $-$ the string tension $\s$ $-$ is proportional
to the quadratic Casimir of the group representation.
\item {\bf N-ality Dependence.}  Asymptotically, the force between charged fields in an SU(N) gauge theory depends 
only on the so-called ``N-ality" of the group representation, given by the number of boxes mod $N$ in the
Young tableau of the representation.
\end{enumerate}
Both of these dependencies have been observed in numerical simulations \cite{Bali1,dFK}; there are also rather
convincing arguments, based on energetics, for N-ality dependence at large distances.  

     We have already mentioned that the color electric field between quarks is collimated
into a flux tube; this precludes long-range van der Waals forces or color dipole fields.   
In addition, string theory models of hadrons predict 
a universal, coupling-constant independent correction to the static quark
potential \cite{Luscher}
\beq
               V_{string}(R) = - {\pi\over 12 R} \; .
\label{LuscherTerm}
\eeq
There is evidence, again from numerical simulations, for the existence of this small correction to the 
linear potential, as well as a spectrum of excitations of the confining electric 
flux tube \cite{Kuti}. 
  
   Taken together, these features of the static quark potential are quite restrictive;  a
completely satisfactory theory of confinement should account for all of them.

\goodbreak

\section{Theories of Quark Confinement}

   There are a number of different approaches to the quark confinement problem.
Probably the most popular idea is that the QCD functional integral is dominated by some
special class of field configurations which cause the expectation value of a large Wilson loop to fall off 
exponentially with the minimal area of the loop, i.e.\ $W(C)\sim \exp[-\s A(C)]$.  For large rectangular loops,
this behavior implies a linear static quark potential, with string tension $\s$.    The leading candidates for
these special configurations are magnetic monopoles and center vortices, although other objects such
as merons \cite{merons}, and calorons \cite{KvB,LL} have also been advocated.
A different approach is based on the special properties of quantization in Coulomb gauge, as we will describe
briefly below.  Another idea is to try to solve non-perturbatively for quark and gluon propagators and vertex
functions, analytically by an infrared expansion of the complete set of Schwinger-Dyson equations, and
numerically 
by solving a truncated set of these equations.  Finally, there is a fascinating relationship
between gauge theory in $D=4$ dimensions and string theory quantized in a special ten-dimension background 
geometry known as anti-DeSitter space.  This is the AdS-CFT correspondence

       Each of these ideas have been the subject of intense study, and the most we can do here is to
give a brief indication of what they are all about. For some of these scenarios, surprising and
interesting relations between them have been discovered, and some of those will be mentioned
along the way.  For a more detailed discussion of material in sections 3.1-3.3 below, see the review
article by one of us \cite{review}.

\subsection{Magnetic Monopoles and Dual Superconductivity}

       The linear static potential would be explained if we could understand why the color electric
field, between a quark and an antiquark, should be collimated into a cylindrical region $-$ a flux tube $-$ of
constant or nearly constant cross-section.   There is a very suggestive example in low temperature physics
known as the Meissner effect:  magnetic fields in type II superconductors are in fact collimated into magnetic
flux tubes, known as Abrikosov vortices.  If magnetic monopoles existed in nature, and a monopole-antimonopole
pair were placed in a type II superconductor, the monopoles would be connected by a magnetic flux tube,
and energy stored in the magnetic field would grow linearly with monopole separation.  This example led 
't~Hooft \cite{thooft1} and Mandelstam \cite{Mandelstam} to suggest that the QCD vacuum is a ``dual superconductor",
the word ``dual" in this case meaning an interchange in the roles of the electric and magnetic fields.  Instead of
magnetic charge confined by a magnetic flux tube in a condensate of electrically charged objects (Cooper pairs),
the idea is that color electrically charged objects (quarks) are confined by an electric flux tube in a condensate
of magnetically charged objects (magnetic monopoles).

     The identification of magnetic monopoles in a non-abelian gauge theory requires the selection of an abelian 
subgroup of the gauge group.   In a theory with a Higgs field $\phi$ in the adjoint representation of the gauge
group, such as the Georgi-Glashow model, an expectation value $\langle \phi \rangle \ne 0$ breaks the SU(N)
symmetry to an abelian $U(1)^{N-1}$ subgroup.\footnote{Actually, a local gauge symmetry cannot be 
spontaneously broken \cite{Elitzur}, and 
the distinction between the ``broken" or ``Higgs" phase, and the unbroken phase, is rather subtle.  But the 
``broken gauge symmetry" terminology is common, even if not strictly correct, and we will continue to use that 
terminology here.}
By fixing to a unitary gauge, so that $\langle \phi \rangle$ has some definite direction in the Lie algebra, gauge
transformations in the abelian subgroup leave this direction unchanged.  Magnetic monopoles can then be 
identified from the abelian magnetic field associated with the
gauge bosons of the abelian subgroup.   In QCD there is no Higgs field, but 't Hooft \cite{thooft2} proposed
that a composite gluonic operator, transforming like a Higgs field in the adjoint representation of
the gauge group, could also serve the purpose of singling out an abelian subgroup.

     Numerical studies of the monopole mechanism have gone in two directions.  The first, pioneered by the
Kanazawa group \cite{Suzuki}, emphasizes a particular composite operator, or, equivalently, a particular gauge which leaves
a remaining $U(1)^{N-1}$ gauge symmetry.  This gauge is the maximal abelian gauge, and on the lattice it is the
gauge which makes link variables as diagonal as possible; {\it e.g.\/} in SU(2) gauge theory, the object is to
maximize
\beq
             R = \sum_{x,\m} \mbox{Tr}[U_{\m}(x)\s_{3}U_{\m}^{\dg}(x)\s_{3}]
\eeq
where $\s_{a}$ denote the Pauli matrices.  The lattice link variables, which take values in the
full SU(N) group, are then projected to the $U(1)^{N-1}$ subroup; this procedure is known as ``abelian projection".
One can then identify monopole worldlines in the abelian projected lattice, and check to see if this monopole content
gives the correct string tension and other static properties of QCD.  The idea is successful in a number of respects,
the main difficulty is representation dependence:  the force between quarks depends on their $U(1)^{N-1}$ electric 
charges, rather than their N-ality \cite{j3}. 

    Another approach, that has been developed largely by the Pisa group \cite{Pisa}, is to define a monopole creation operator
$\phi^{M}(x)$ in SU(N) lattice gauge theory, with the monopoles again defined in some gauge, and check to see
that $\langle \phi^{M} \rangle \ne 0$.  In an ordinary abelian Higgs theory, a vacuum expectation value 
$\langle \phi \rangle \ne 0$ breaks the U(1) gauge symmetry of the theory, and we have an electric 
superconductor.  The idea is that in a non-abelian gauge theory with no elementary Higgs fields, an expectation value
$\langle \phi^{M} \rangle \ne 0$ implies the breaking of a dual magnetic symmetry, and confining gauge theories
exist in the ``dual superconductor" phase.     

     The idea of dual superconductivity received a great boost from the work of 
Seiberg and Witten \cite{SW}, who were able to show analytically that in certain supersymmetric gauge theories, monopole condensation actually does take place.  In these
theories, unlike QCD, there exists an elementary Higgs field which can be used to single out
a unique abelian subgroup; fixing the abelian subgroup by a composite operator is unnecessary.   In these particular supersymmetric
theories, it is possible to derive explicitly an effective dual abelian Higgs action, at least if the effects of gluons not belonging to the abelian subgroup are ignored.  But in the resulting effective theory,  as in other 
``monopole dominance" models \cite{j3}, the asymptotic string tension between quarks of a given abelian charge depends only on that
abelian charge, and not on the quadratic Casimir or the N-ality of the associated SU(N) representation.   Related to this fact is a certain multiplicity of Regge trajectories  \cite{Strassler}, which is also rather unlike QCD.
 
     Monopoles also arise in investigations of objects known as ``calorons", which are instantons at finite
temperatures.  Instantons are semi-classical solutions of the Euclidean-time gauge field equations, and finite
temperature corresponds to a finite, periodic extension in the time direction.  Recent studies 
\cite{Gattringer:2003uq} center on a type
of caloron solution with non-trivial holonomy found by Kraan and van Baal \cite{KvB}, and by Lee and Lu \cite{LL}.  The calorons can
be thought of as bound states of monopoles, which tend to move apart from one
another as the temperature is lowered \cite{Pierre}.  It has been suggested that confinement could be attributed to
caloron dynamics. This again raises the question of how a caloron-based confinement mechanism 
would obtain the correct N-ality dependence of the string tension extracted from Wilson loops (see, however, ref.\ \cite{Ilgenfritz}). 

\subsection{Center Vortices}

    We have already mentioned that the asymptotic string tension depends only on the N-ality of the quark charge.
This fact is important, and easily understood.  First of all, there are an infinite number of SU(N) representations, but
only a finite number of representations of the $Z_{N}$ subgroup, so the representations of SU(N) can be divided into
classes, each with the same N-ality.  Gluons have N-ality equal to zero.  This means that when gluons bind to a
color charge in some representation $r$, the resulting bound state is in a color representation $r'$ with the same
N-ality as $r$.  So as a quark in representation $r$ separates from its antiquark, gluons can be pair-created out
of the vacuum, and bind to the quark and antiquark, reducing the color charge of each.  However, the color
charge can only be reduced to the lowest dimensional representation with the same N-ality as $r$.  For example,
in SU(2) gauge theory the center group is $Z_{2}$, and representations can be divided into $j=$half-integer, with
N-ality one, and $j=$integer, with N-ality zero.  As two quarks in the $j=3/2$ representation separate, pair-created
gluons can bind to the quarks reducing the color charge to $j=1/2$.  So the asymptotic string tension of
heavy $j=3/2$ quarks is the same as that of $j=1/2$ quarks, and in fact, by the same argument, the asymptotic
string tension of all N-ality=1 quarks are the same.  Likewise, two quarks in the adjoint ($j=1$) representation can
bind to gluons, forming two color singlets.  The asymptotic string tension of $j=1$ quarks is zero, as is the string
tension of quarks in any N-ality=0 representation.

      It is helpful to think about this N-ality dependence in the context of the Euclidean functional integral over
field configurations.   In a Monte Carlo simulation, Wilson loops are simply averaged over some finite set of
lattice configurations, generated stochastically with the appropriate probability weighting.  How do these configurations
manage to produce asymptotic string tensions that depend solely on the N-ality of the group representation of the
Wilson loop?

     The answer to this question comes from an interesting direction.  In 1978 't Hooft \cite{thooft3} introduced a loop operator
$B(C)$ in SU(N) gauge theories, intended to be in some sense ``dual" to the ordinary Wilson loop operator $W(C)$, 
with the rather beautiful commutation property
\beq
             B(C) W(C') = e^{2\pi i/N} W(C') B(C)
\eeq
if curves $C$ and $C'$ are topologically linked to one another.  't Hooft argued, just from this commutation relation
and the presumed existence of a mass gap,
that $\langle B(C) \rangle$ has a perimeter-law falloff $\exp(-\m P(C))$ in the confined phase of
a gauge theory, and an area-law falloff in the non-confining Higgs phase.  This is precisely opposite to the
behavior of Wilson loops in those two phases.  It turns out that $B(C)$ is the creation operator for an object known
as a ``center vortex".  Roughly speaking, a center vortex is a tube of magnetic flux, in which the exponential
of the vortex flux, measured by a Wilson loop $P\exp[i\oint_{C'}dx^\m A_{\m}]$ running around the vortex, 
takes values in the $Z_{N}$ center of the SU(N) gauge group.  More precisely, creation of a center vortex which is
topologically linked to a given Wilson loop changes the Wilson loop by a multiplicative factor equal to a center element.
A Wilson loop in some group representation $r$ is multiplied by a factor $\exp[2\pi i k/N] \in Z_{N}$ which depends
only on the N-ality $k$ of the representation $r$.   This is the crucial property.  In order to have a confinement mechanism
in which the string tension of a Wilson loop depends only on the N-ality of the loop, it is necessary to have configurations
which affect loops of the same N-ality in the same way.  Center vortices are the only known field configurations which satisfy
this requirement.   In D=4 Euclidean dimensions these objects are actually surfaces; they may be thought of as having been swept out
by a magnetic vortex loop as it propagates in time. In D=3 dimensions loops can be topologically linked to other loops;
in D=4 dimensions loops link to surfaces.

    The center vortex confinement mechanism, as elaborated in refs. \cite{thooft3,others} is quite simple:  Center vortices percolate
throughout the vacuum, and Wilson loops derive an area law from random fluctuations in the number of vortices piercing the loop.
This proposal lay dormant for about fifteen years, but was revived after a series of numerical investigations \cite{FGO,Tubingen,dFE}
turned up rather strong evidence in its favor.  This evidence is reviewed in detail in ref.\ \cite{review}.  Briefly, the numerical techniques
are very close to those employed in abelian projection.  Lattice configurations are fixed to a gauge which leaves a residual
$Z_{N}$ invariance, and SU(N) group-valued link variables are then projected to the nearest element of the $Z_{N}$ subgroup.
The excitations of the projected configurations are thin center vortex configurations known as ``P-vortices", and these thin 
vortex sheets appear to lie within thick vortex  surfaces in the unprojected gauge theory.  Although gauge fixing is used
in the identification, P-vortices correlate with both the gauge-invariant
action density and gauge-invariant Wilson loops on the unprojected lattice.  P-vortex areas scale with lattice
coupling as expected from asymptotic freedom, and by themselves produce an area law falloff for Wilson loops with roughly
the right string tension.  When vortices are removed from the original lattice, the string tension drops to zero, topological
charge vanishes, and chiral symmetry is unbroken.

    It is worth expanding a little on vortex removal.  Let $U(C) \in$ SU(N) be a Wilson loop around curve $C$, and let $Z(C) \in Z_{N}$
be the value of the Wilson loop in the projected configuration.  In SU(2) gauge theory, $Z(C) = (-1)^{l(C)}$, where $l(C)$ is
the number of P-vortices linked to the loop.  Denote the corresponding Wilson loop in the vortex-removed configuration
by $U'(C)$.   These expressions have a simple relationship to one another
\beq
             U'(C) = Z(C) U(C) = (-1)^{l(C)} U(C) .
\label{vrem}
\eeq
Numerically, it is found that the projected and unprojected Wilson loop expectation values, 
$\langle Z(C)\rangle$ and $\langle \mbox{Tr}[U(C)]\rangle$
respectively, both have area-law falloffs with approximately the same string tension, while the string tension of Wilson loops
in the vortex-removed configuration $\langle \mbox{Tr}[U'(C)]\rangle$ is vanishing.   The only way that this can happen,
in view of \rf{vrem}, is that the fluctuations in the sign of $\langle \mbox{Tr}[U(C)]\rangle$ are correlated to fluctuations
in P-vortex linking number.   This correlation of the linking number of P-vortices with the sign of gauge-invariant Wilson loops, while 
certainly not a proof of the vortex confinement mechanism, argues strongly in its favor.
 
    It is also found that the monopole worldlines of the abelian projected lattice lie on P-vortex worldsheets \cite{j3}.   A center vortex
can, in fact, be thought at any given time as a kind of monopole-antimonopole chain, in which the abelian magnetic flux 
of the monopoles and antimonopoles is collimated along the vortex line.  This means that the vortex and abelian monopole
pictures are not really antagonistic; the collimation of the monopole flux is exactly what is required for the monopole
picture to satisfy the required N-ality dependence for the asymptotic string tension.   We also note that
Casimir scaling, in the vortex picture, is due to the finite thickness of center vortices \cite{Cas}, and spatial variations 
of flux within the vortex core \cite{g2}.   In 
a gauge theory based on the group G(2), which has a trivial center subgroup,  the prediction of the vortex theory 
is that the asymptotic string tension is zero.   This agrees with expectations, since in G(2) gauge theory gluons can combine
with quark charges in any representation to form a color singlet.
    
    The main reservations to the vortex picture are numerical:  the string tensions in the projected lattices are not quite the
same as for the unprojected lattice, and there are concerns regarding the gauge-fixing procedure, which is plagued by
Gribov copies \cite{Bornyakov}.

\subsection{Coulomb Energy and the Gribov Horizon}

    By definition gauge invariance implies a redundancy in the degrees of freedom of a gauge field.
Hamiltonian dynamics requires an elimination of this redundancy via a gauge choice, resulting
in a formulation involving the correct number of physical degrees of freedom.  In Coulomb gauge,
in particular, there is a very suggestive separation between the electric energy due to the
longitudinal ({\it i.e.\/} Coulombic) and transverse electric fields.  Classically, the Coulomb gauge
Hamiltonian has the form\footnote{Quantum-mechanically there are some operator-ordering modifications,
which we will not discuss here.}
\bea
H %&=& 
=\oh \int d^3x ~ (\vec{E}^{a,tr} \cdot \vec{E}^{a,tr}
 + \vec{B}^a \cdot \vec{B}^a)
%\non \\
%  & & \qquad 
+ \oh \int d^3x d^3y ~ \r^a(x) K^{ab}(x,y) \r^b(y)
\eea
where $\r^a$ is the (matter plus gauge field) color charge density,
$\vec{E}^{a,tr}$ is the transverse color electric field
operator, and
\begin{equation}
       K^{ab}(x,y) = \left[ M^{-1}(-\nabla^2)M^{-1}
                      \right]_{x,y}^{ab}
\label{Dansway}
\end{equation}
is the instantaneous Coulomb propagator.  The Coulomb interaction energy is
given by the non-local term in the Hamiltonian, involving $\rho K \rho$. 
In an abelian theory,  $K(x,y)$ is simply proportional to $1/|\vec{x}-\vec{y}|$.  
In a non-abelian theory, $K^{ab}(x,y)$
is dependent on the gauge-field through the Faddeev-Popov operator
\beq
         M^{ac} = - \pa_i D_i^{ac}(A)
                = -\nabla^2 \d^{ac} - \e^{abc} A_i^b \pa_i \; .
\eeq
Could it be that the vacuum expectation value of the Coulomb energy, for static sources,
leads to an asymptotically linear, rather than $1/r$, potential?  

    Gribov \cite{Gribov} and Zwanziger \cite{Dan1} have argued in favor of this possibility.
The argument is roughly as follows:  In Coulomb gauge, the integration over gauge fields
needs be restricted to $A$-fields satisfying $\nabla \cdot A = 0$, \emph{and} for which the Faddeev-Popov
operator is positive; {\it i.e.\/} for which the eigenvalues of the $M$ operator are all positive.  This positivity
condition restricts the gauge fields to a subspace of gauge-field configuration space; the boundary of
this region, where the $M$ operator develops a zero eigenvalue, is known as the ``Gribov horizon".
Since the Coulomb propagator $K^{ab}(x,y)$ depends on the inverse of $M$ this operator becomes very
large in the neighborhood of the Gribov horizon.  Now, since the dimension of configuration space is very large, it is
reasonable that the bulk of configurations are located close to the horizon
(just as the volume measure $r^{d-1} dr$
of a ball in $d$-dimensions is sharply peaked near the radius of the
ball).  Since it is the inverse of the $M$ operator which appears in
the Coulomb energy, it is possible that the near-zero eigenvalues of
this operator will enhance the magnitude of the energy at large quark
separations, possibly resulting in a confining potential at large
distances.   Moreover, it is possible to prove the following inequality \cite{Dan2}:  If $V(R)$
is the static quark potential ({\it i.e.\/} the minimal energy of a physical state containing
two static quark-antiquark sources in the fundamental representation), and $V_{coul}(R)$ is
the Coulomb potential obtained from the vacuum expectation value $\langle \rho K \rho \rangle$, then
\beq
          V(R) \le V_{coul}(R) .
\eeq
This means that ``Coulomb confinement" is a necessary (but not sufficient) condition for
confinement.   The interesting question is whether $V_{coul}(R)$ is linear and, if linear,
whether the corresponding string tension $\s_{coul}$ equals the string tension of the static
quark potential.

    These questions have been investigated numerically, via lattice Monte Carlo in Coulomb
gauge.  The answer is that $V_{coul}(R)$ does indeed rise linearly \cite{GO1,Nakamura:2005ux}.  
Moreover, the infrared divergent Coulombic energy of an isolated charge comes about by precisely
the mechanism suggested by Gribov and Zwanziger: a large density of eigenvalues of the Faddeev-Popov
operator in the neighborhood of the zero eigenvalue \cite{GOZ}.  When center vortices are removed
from lattice configurations, the Coulombic energy is non-confining, and the Faddeev-Popov
eigenvalue distribution resembles that of the abelian theory. Together with the fact that center vortices
are field configurations lying on the Gribov horizon, this suggests a close relationship between
these two confinement scenarios.

    On the other hand, it also turns out that the Coulomb string tension is about three times larger
than the string tension of the static quark potential, so the behavior of $K(x,y)$ cannot be the
whole story behind confinement.  There is, in fact, an even more basic objection:  Any theory of
confinement based on one-gluon (or one quasi-particle) exchange will have difficulties in explaining
why the color electric field is collimated into a flux tube.  In general, one-particle exchange models
of the confining force give rise to long-range dipole fields.  If a proton, say, were held together
by one-gluon exchange forces, it is hard to see why there could not be long-range color van-der-Waals
forces among distant protons, contrary to observation.

    One interesting approach, followed in ref.\ \cite{Adam},   
is to try to construct physical states in Coulomb gauge, whose energy is lower than that of a
quark-antiquark pair plus their Coulomb field, by adding constituent gluons. Possibly this could also
help with the problem of the long-range dipole field.  We might imagine
that as a quark and antiquark separate, they pull out between them a ``chain" of constituent gluons,
with each gluon in the chain bound to its nearest neighbors by Coulombic forces.  This is the
``gluon chain model" \cite{gchain}, and it provides a picture of the QCD flux tube as a kind of
discretized string.  Whether this gluon-chain picture will eventually emerge from investigations 
in Coulomb gauge or, alternatively, from a recent worldsheet formulation of gauge theory
quantized in light-cone gauge \cite{Charles}, remains to be seen.

      Before leaving this topic, we might ask: Given that instantaneous one-gluon exchange results in a 
linear attractive potential between
a quark and an antiquark, what would be situation for two quarks?  Would we end up with a finite
energy color non-singlet state and a linear repulsive potential?  That would, of course, be a real disaster for this
approach.   In fact, in calculating Coulomb interaction energies of composite states one has to carefully take into 
account the cancellation of divergences which are
encountered in both the quark self-energy and one-gluon exchange terms.  It turns out that for color singlets, these divergences
precisely cancel, leaving a finite attractive potential, while in non-singlets the divergent self-energies are not cancelled,
and the energy of the non-singlet state is infinite.   This cancellation is discussed in ref.\ \cite{GOZa}, and demonstrated, in the
context of a Bethe-Salpeter approach, in ref.\ \cite{Alkofer:2005ug} and references therein.

\subsection{Functional Approaches}

Functional approaches employed to the infrared behaviour of QCD are Schwinger-Dyson Equations (SDEs)
and Renormalization Group Equations\footnote{A combination of both methods has recently allowed to 
{\em uniquely} determine the infrared behavior of all Green functions of Landau gauge Yang-Mills
theory \cite{Fischer:2006}.}, for a recent review see \cite{Fischer:2006ub}. In the Landau gauge the
analytical treatment of these equations in the far infrared have provided a number of exact
inequalities for the infrared exponents  of gluon and ghost one-particle irreducible (1PI) Green
functions.  Gauge fixing is hereby performed by the standard Faddeev-Popov method supplemented by
auxiliary conditions such that the generating functional consists of an integral over gauge field
configurations that are contained in the first Gribov region. The employed method has been justified
using ghost-free stochastic quantisation   \cite{Zwanziger:2003cf}. The resulting SDEs for 
1PI-Green functions have been solved analytically in the infrared to all orders in a skeleton
expansion ({\it i.e.\/} a loop expansion using full  propagators and
vertices)~\cite{Alkofer:2004it}. It turns out that these Green's  functions are infrared singular in
case {\em all} external momenta go to zero.  A remarkable property of the infrared solution is the
fact that it is generated by exactly those parts of the SDEs that involve ghost loops. In other
words: the Faddeev-Popov determinant dominates the infrared behaviour of  non-Abelian Yang-Mills
theories. Thus an infrared asymptotic theory can be obtained by `quenching` the Yang-Mills action,
{\it i.e.\/} setting $\exp(-S_{YM})=1$ in the generating functional \cite{Zwanziger:2003cf}. The
solution of this asymptotic theory is given  by  power laws. 

The basic examples for power law solutions are
the ghost and gluon propagators 
\beq 
D^G(p^2) = -\frac{G(p^2)}{p^2} \, , \ \
D_{\mu \nu}(p^2)  = \left(\delta_{\mu \nu} -\frac{p_\mu
p_\nu}{p^2}\right) \frac{Z(p^2)}{p^2} \, . 
\eeq
The corresponding power laws in the infrared are
\beq
G(p^2) \sim (p^2)^{-\kappa}, \hspace*{1cm} Z(p^2) \sim (p^2)^{2\kappa}\,.
\label{kappa}
\eeq
Since $\kappa$ is positive \cite{Watson:2001yv} one obtains an infrared enhanced ghost and an 
infrared suppressed 
gluon propagator. In Landau gauge an explicit value for 
$\kappa$ can be derived from the observation that the dressed ghost-gluon vertex
becomes (almost) bare in the infrared, one then obtains $\kappa = (93 - \sqrt{1201})/98 \approx 0.595$
\cite{Lerche:2002ep,Zwanziger:2001kw}.\footnote{Dynamical quarks do not change the infrared behavior
of the gluon and ghost propagators \cite{Fischer:2003rp}.} 

\setcounter{footnote}{0}

Let us shortly digress here and mention in which sense this leads to the so-called ``kinematic
confinement'' of transverse gluons.\footnote{The mechanism becomes most transparent in a covariant
formulation which includes the choice of a covariant gauge, of course. However, the arguments for  the
positivity violation in the propagator of transverse gluons are analogously applicable in the Coulomb
gauge, and numerical evidence for it is equally firm as in the Landau gauge case, see {\it e.g.\/} refs.\ 
\cite{Dan1,Cucchieri:2000gu,Szczepaniak:2003ve,Feuchter:2004mk,Nakamura:2005ux}.}  First we note that
covariant quantum theories of gauge fields require indefinite metric spaces. Abandoning the positivity of
the representation space already implies to give up one of the axioms of standard quantum field theory.
Maintaining the much stronger principle of locality gluon confinement then naturally relates to the
violation of positivity in the gauge field sector, see {\it e.g.\/} ref.\ \cite{Alkofer:2000wg}. Similar to
QED, where the Gupta-Bleuler prescription \cite{Bleuler50} is to enforce the Lorentz condition on physical
states, a semi-definite physical subspace can be defined as the kernel of an operator. The physical states
then correspond to (equivalence classes of) states in this subspace.  Covariance implies, besides
transverse photons, the existence of longitudinal and timelike (``scalar'') photons in QED. The latter two
form metric partners in the indefinite space: They cancel against each other in every $S$-matrix element
and therefore do not contribute to observables. 

In QCD cancelations of unphysical degrees of freedom in the $S$-matrix  also occur but are more complicated
due to the self-interaction of the gluons. A consistent quantum formulation in a functional integral
approach leads to the introduction of ghost fields \cite{Faddeev67}. The proof of the cancelation of
longitudinal and timelike gluons in every $S$-matrix element to all orders in perturbation theory has been
possible by employing the BRST symmetry~\cite{Becchi:1976nq} of the covariantly gauge fixed theory. At this
point one has achieved a consistent quantization.

Based on the BRST formalism and implications of the Gribov horizon \cite{Zwanziger:1991ac} 
positivity violation of the propagator of transverse gluons has been a
long-standing conjecture for which there is now compelling evidence, see   {\it
e.g.\/}~ref.~\cite{Alkofer:2003jj} and references therein. The basic features underlying these gluon
properties, namely the infrared suppression of correlations of transverse gluons and the infrared
enhancement of ghost correlations, has been verified in quite a number of lattice Monte-Carlo
calculations and different functional approaches. One then concludes that transverse gluons possess
metric partners, they form a so-called BRST quartet together with gluon-ghost, gluon-antighost and
gluon-ghost-antighost states. Gluon confinement then occurs as necessarily complete cancelation between
amplitudes (Feynman diagrams) containing  these states as asymptotic states (as external lines).  
This is in line with the na\"ive interpretation of a gluon propagator which vanishes in the infrared:
A zero in the propagator at $p^2=0$ implies that there is no propagation of gluons at long
distances.\footnote{For chromomagnetic gluons this picture persists in the high-temperature phase
of QCD \cite{Maas:2005hs}.}

An additional important consequence of this infrared solution for gluons and ghosts is the
qualitative universality of the running coupling in the infrared.
Renormalisation group invariant couplings can be defined from either of the
primitively divergent vertices of Yang-Mills-theory, {\it i.e.\/} from the
ghost-gluon vertex, the three-gluon vertex, or the four-gluon
vertex. All three couplings approach a fixed point in the infrared. However,
the explicit value of the fixed point may be different for each coupling. For a bare 
ghost-gluon vertex one obtains $\alpha^{gh-gl}(0) \approx 8.92/N_c$ \cite{Lerche:2002ep,Alkofer:2002ne}; 
the other couplings have not been determined yet.\footnote{For all these couplings the infrared fixed point
behaves like $1/N_c$ thus obeying the correct large-$N_c$ behavior for all values of $N_c$.
A detailed investigation of the large-$N_c$ limit of this approach is, however, still lacking.} 
This behavior  sheds light on the existence of the power law
solutions: pure Yang-Mills theory  becomes approximately conformal in the far
infrared.

As explained in the introduction the static quark potential is a property of would-be infinitely heavy
quarks. To extend the infrared analysis to full QCD \cite{Alkofer:2006gz} one concentrates first on the
quark sector of quenched QCD and chooses the masses of the valence quarks to  be large, {\it i.e.\/}  $m
> \Lambda_{\tt QCD}$.   The remaining scales below $\Lambda_{\tt QCD}$ are those of the external
momenta of the Green functions. Without loss of generality these can be  chosen to be equal, since
infrared singularities in the corresponding loop integrals  appear only when all external scales go to
zero \cite{Alkofer:2004it}.   One can then employ SDEs to determine the selfconsistent solutions  in
terms of powers of the small external momentum scale $p^2 \ll \Lambda_{\tt QCD}$. The SDEs which have to
be considered in addition to the SDEs of Yang-Mills theory are the one for the quark propagator and the
quark-gluon vertex. The dressed quark-gluon vertex $\Gamma_\mu$ consists in general of twelve Dirac
tensor  structures. Some of these tensor structures would have to vanish if chiral symmetry would not be
broken (either explicitely or dynamically). Especially those Dirac-scalar structures are, in the
chiral limit, generated non-perturbatively together with the dynamical quark mass function in a
self-consistent fashion: Dynamical chiral symmetry breaking reflects itself thus not only in the
propagator but also in a three-point function.

An infrared analysis of the full set of DSEs reveals that a solution exists  such that vector and {\em scalar}
components of the quark-gluon vertex are infrared divergent with an exponent related to $\kappa$
\cite{Alkofer:2006gz}. A
numerical solution of truncated set of SDEs confirms this infrared behavior. Similar to the Yang-Mills
sector it is the diagram containing the ghost loop that dominates.  Thus all effects from the Yang-Mills
sector are generated by the infrared asymptotic theory described above. More importantly, in the quark
sector the driving pieces of this solution is the scalar Dirac amplitude of the quark-gluon vertex and
the scalar part of the quark propagator. Both pieces are only present when chiral symmetry is broken.
%\footnote{This is corroborated by recent lattice calculations \cite{Gattringer:2006ci} 
%where a relation between the spectral properties 
%of the Dirac operator and confinement has been noticed.}

\begin{figure}
\centering
\includegraphics[width=12cm]{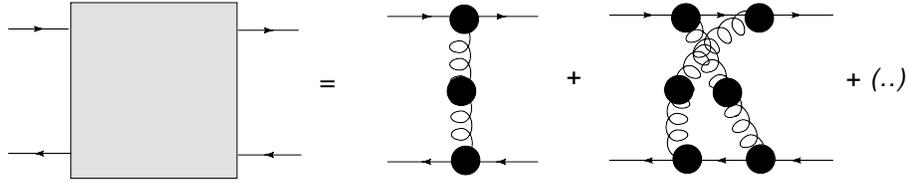} 
\caption{The four-quark 1PI Green's function and the first terms of its skeleton expansion,
adapted from ref.\ \cite{Alkofer:2004it}.}
\label{qq-1PI}
\end{figure}
The static quark potential is obtained from the four-quark 1PI Green's
function $H(p)$, which is given in Fig.~\ref{qq-1PI} together with its skeleton expansion. From
the infrared analysis one infers that $H(p) \sim (p^2)^{-2}$ in the infrared.
From the usual relation
\beq
V({\bf r}) = \int \frac{d^3p}{(2\pi)^3}  H(p^0=0,{\bf p})  e^{i {\bf p r}} 
\ \ \sim \ \ |{\bf r} |
\eeq
between the static four-quark function $H(p^0=0,{\bf p})$ and the quark potential $V({\bf r})$
one therefore obtains a linear rising potential. Correspondingly, the running coupling from the 
quark-gluon vertex turns out to be proportional to  $1/p^2$
in the infrared, {\it i.e.\/} contrary to the couplings  from the Yang-Mills vertices this
coupling is  singular in the infrared.

Already the first term in the skeleton expansion, {\it i.e.\/} the effective,  nonperturbative
one-gluon exchange displayed in Fig.~\ref{qq-1PI}, generates this result. Since the following terms
in the expansion are equally enhanced in the infrared, the string tension will be built up by
summing over an infinite number of diagrams. The latter property is bad news for the usefulness 
of the approach but it had to be expected in the first place.
Since already an effective, nonperturbative one-gluon exchange generates the confining
potential one is again, as in the previous subsection, confronted with the problem
of unwanted van-der-Waals forces. The suppressed gluon propagator looks at first sight helpful
because it implies that there are no long-range correlations between the gluons, {\it i.e.\/} the
gauge fields, and thus for chromoelctric and chromomagnetic fields at large
distances. However, the problem of avoiding long-range multipole fields 
has only be shifted from the two-point correlation to a specific three-point
function, namely the quark-gluon vertex.
In addition, as very likely every picture based on a finite number of quasi-particles has to fail in
explaining  the L\"uscher term (\ref{LuscherTerm})  one can already conclude that the series in
Fig.~\ref{qq-1PI} needs to be an infinite one if the picture were able to describe quark confinement
correctly.

Last but not least, N-ality can occur in such pictures only if cancelations as {\it
e.g.\/}  between gluons and adjoint quarks will take place. Casimir scaling, on the other hand,
requires that at intermediate distances these cancelations are still absent or incomplete.
Explaining these features of confinement is still a completely unsolved challenge within functional
approaches.

\subsection{AdS/CFT correspondence}

There is compelling evidence that a type-IIB closed superstring theory in ten dimensions is dual to
an ${\cal N}=4$ super-Yang-Mills (sYM) theory. The space-time in the superstring theory  is such
that five dimensions form a sphere, and the other five dimensions a non-compact anti-de Sitter
space, briefly denoted by AdS$_5$. Hereby the sphere has a (positive) radius $R$ and AdS$_5$ a
negative curvature of the same scale
\beq
ds^2= R^2 d\Omega_5 + \frac{R^2}{r^2} dr^2 + \frac{r^2}{R^2} \eta _{\mu\nu} dx^\mu dx^\nu .
\eeq 
The ${\cal N}=4$ sYM theory has as much supersymmetry a gauge
theory can have: It is a conformal field theory (CFT). Mathematically, the duality is built on the
fact that the isometry group of AdS$_5$ is isomorphic to the conformal group of four-dimensional
Minkowski space, SO(4,2).\footnote{It is an irony of this field that this first example of AdS/CFT
duality does not confine because ${\cal N}=4$ sYM theory is exactly conformal. When a large Wilson
loop is introduced on the boundary of AdS$_5$ the red shift factor $r^2/R^2$
allows the minimal surface
spanning the loop to stay finite implying a perimeter instead of an area law for the sYM 
Wilson loop.}
Employing a suitable background metric the gauge coupling is inversely proportional to the string
tension:
\beq
g^2N_c = R^4/{\alpha '}^2.
\eeq
This implies that the strong coupling gauge theory is dual to weak coupling string theory, and thus,
as the low-energy limit of superstring theory is supergravity, to weak coupling gravity.

Real-world QCD is not supersymmetric. Therefore one needs to break supersymmetry and therefore also
conformal invariance. The corresponding models typically modify the metric in the infrared by
introducing cutoffs, or equivalently black hole type backgrounds. 
The corresponding minimal distance is then identified with the inverse of the
QCD scale:\footnote{In most corresponding calculations the strong-coupling limit of the smallest 
glueball mass is used to set the scale.}
\beq
r_{\rm min} = 1 /\Lambda _{\tt QCD}.
\eeq
In those black hole metrics the minimal surface spanning a Wilson loop of increasing size eventually
has to approach $r=r_{\rm min}$. Beyond this point no red shift factor contributes to 
the area of the surface, it grows proportional to $R^2r^2_{\rm min}$ providing a non-vanishing string
tension 
\beq
\sigma = 1 / 2\pi {\alpha '}_{\tt QCD} =  R^2r^2_{\rm min}/ 2\pi {\alpha '} = \sqrt{2g^2N_c} / 2\pi
\Lambda^2 _{\tt QCD}. 
\eeq  
Another approach is to introduce so-called ``fractional D-branes'' to break the 
supersymmetry and conformal invariance, {\it c.f.\/} ref.\ 
\cite{Klebanov:2000hb}.

Since the first considerations of Wilson loops within AdS/CFT correspondence, see {\it e.g.\/} ref.\
\cite{Maldacena:1998im}, a large number of papers appeared on the subject, and a
summary of these
developments is far beyond the
scope of the present article.   We nevertheless want to note that the N-ality condition
has recently been shown to be fulfilled in this approach \cite{Armoni:2006ri}.

Phenomenogical tests of AdS/CFT correspondence are abundant, it has especially been successful in
reproducing general properties of scattering processes of QCD bound states. Hereby confinement can
be simulated by cutting off the extension of hadron wave function into the ``fifth''dimension
\cite{Polchinski:2001tt}. The interested reader can obtain a first impression from refs.\
\cite{Brodsky:2005en}, the references therein provide a reasonable guide for further reading.

It is plainly obvious but it should nevertheless be emphasized here: The AdS/CFT correspondence
provides no explanation for confinement. It is a calculational tool relating low-energy, 
non-perturbative QCD to weak-coupling gravity where the background has been {\em chosen} such
to provide confinement in QCD. However, as some of the related problems
can be treated much easier in the gravity language the approach has and likely will furthermore
provide insights into the special properties of possible confinement scenarios
in QCD.

\section{Conclusions}

     It is odd to have a complete theory of one of the four well-established forces of nature $-$
the strong nuclear force $-$ and still not have general agreement, after more than thirty years of effort, 
on how that force really works at long distances.   Nevertheless, there \emph{has} been appreciable progress in this
subject, much of it aided by computer simulations.  First of all there is a better appreciation of the general features 
of the confining force, e.g.\ the color group representation dependencies (N-ality, Casimir scaling) of the 
confining potential, and the existence and string-like properties of the color electric flux tube,  which constrain possible 
explanations of confinement. Secondly, there exist a reasonable set of suggestions about the origin of confinement, 
some dating back to the late 1970's and some much more recent, which have, over the last decade, received substantial 
support from lattice Monte Carlo simulations.   It has turned out that a number of these suggestions are related
in interesting ways:  monopole wordlines, essential to dual-superconductor scenarios, are found to lie on center 
vortex worldsheets, and center vortex worldsheets appear to be crucial in some ways to the
confinement scenario in Coulomb gauge.   Both Coulomb and Landau gauge investigations emphasize the
importance of the Faddeev-Popov operator, and the infrared properties of the ghost propagator.   

      There are other proposals for the confinement mechanism which we have not included here.  It is impossible to
provide an exhaustive discussion in a short article, so we have concentrated on those proposals which, in our judgement, 
seem best supported by existing numerical studies or other arguments.   But it is certainly not excluded that progress 
may come from some quite different direction.

   The confinement problem, in our view, is one of the truly fundamental problems in physics.   
Quark confinement is the essential link between the microscopic quark-gluon degrees of freedom of QCD, 
and the actual strong-interaction spectrum of color-neutral mesons, baryons, and nuclei.  Until this phenomenon 
is well understood, something essential is still lacking in our grasp of the foundations of nuclear physics, and the 
deeper mechanisms of non-abelian gauge theory.   Although the confinement problem is hard, the solution is important,
and well worth pursuing.

% Acknowledgements
%
\ack{%
We are grateful to our co-workers, Per Amundsen, Manfried Faber, Christian Fischer, Kurt Langfeld,
Felipe Llanes Estrada, Axel Maas, {\v S}tefan Olejn\'{\i}k, Hugo Reinhardt, Lorenz von Smekal, Charles Thorn,
Pete Watson, and Daniel Zwanziger,  who have shared with us their efforts and their insights into the
confinement problem. Our research is supported in part by the U.S. Department of Energy
under Grant No.\ DE-FG03-92ER40711 (J.G.), and the Deutsche Forschungsgemeinschaft
under Grant No.\ Al279/5-1 (R.A.). }

\vspace{10mm}

%
% References
%

\end{document}